\documentclass[showpacs]{revtex4}
\usepackage{graphicx}
\begin{document}
\title{Energy and entropy conservation for dynamical black holes}
\author{Sean A. Hayward}
\affiliation{Department of Physics, National Central University, Jhongli,
Taoyuan 320, Taiwan\\ {\tt sean\_a\_hayward@yahoo.co.uk}}
\date{revised 28th October 2004}

\begin{abstract}
The Ashtekar-Krishnan energy-balance law for dynamical horizons, expressing the
increase in mass-energy of a general black hole in terms of the infalling
matter and gravitational radiation, is expressed in terms of general trapping
horizons, allowing the inclusion of null (isolated) horizons as well as spatial
(dynamical) horizons. This first law of black-hole dynamics is given in
differential and integral forms, regular in the null limit. An effective
gravitational-radiation energy tensor is obtained, providing measures of both
ingoing and outgoing, transverse and longitudinal gravitational radiation on
and near a black hole. Corresponding energy-tensor forms of the first law
involve a preferred time vector which plays the role for dynamical black holes
which the stationary Killing vector plays for stationary black holes.
Identifying an energy flux, vanishing if and only if the horizon is null,
allows a division into energy-supply and work terms, as in the first law of
thermodynamics. The energy supply can be expressed in terms of area increase
and a newly defined surface gravity, yielding a Gibbs-like equation, with a
similar form to the so-called first law for stationary black holes. A
Clausius-like relation suggests a definition of geometric entropy flux. Taking
entropy as area/4 for dynamical black holes, it is shown that geometric entropy
is conserved: the entropy of the black hole equals the geometric entropy
supplied by the infalling matter and gravitational radiation. The area or
entropy of a dynamical horizon increases by the so-called second law, not
because entropy is produced, but because black holes classically are perfect
absorbers.
\end{abstract}
 \maketitle

\section{Introduction}
Black holes are perhaps the most exotic and energetic objects in the universe.
Their theoretical history is long and winding: the earliest such solution to
the field equations of Einstein's theory of General Relativity \cite{Ein} was
found by Schwarzschild \cite{Sch} almost immediately, but not understood as
such for decades \cite{Kru}. In a few years around 1970, there was rapid
theoretical progress, with the introduction of the term black hole by Wheeler
\cite{Whe} and the development of the classical four laws of black-hole
mechanics \cite{BCH,HE,MTW,Wal}, supposedly analogous to the laws of
thermodynamics. Since then, astrophysical evidence has increasingly accumulated
not only for stellar-mass supernova-remnant black holes, but for supermassive
black holes, mysteriously present at the heart of most if not all galaxies and
powering active galactic nuclei \cite{MBH}. Cataclysmic events such as binary
black-hole mergers are predicted to produce gravitational waves which are
observable on or near our home planet, for which a new generation of detectors
is being developed \cite{LIGO}. Consequently, recent years have seen a great
deal of work on numerical simulations to study how black holes evolve according
to given initial conditions, and what gravitational radiation they may produce
\cite{BBH}.

Such progress leaves the textbook theory of black holes seriously out of date.
Much is known about stationary black holes, for instance the zeroth and first
laws just mentioned, but dynamical black holes are much more complex. Of the
classical laws, only Hawking's area theorem has generality, but it applies to
event horizons, which are theoretical constructs which cannot be located by
mortals. It is quite timely that Hawking has recently recanted, writing that
``a true event horizon never forms, just an apparent horizon'' \cite{GR17}.
Unfortunately, Hawking's definition of apparent horizon \cite{HE} is also not
the most appropriate to define black holes, due to its global nature and
slicing dependence; for instance, the Schwarzschild black hole may be globally
sliced so that there is no apparent horizon \cite{WI}.

About ten years ago, the author began a program to understand local, dynamical
properties of black holes \cite{bhd,bhs}. The basic idea is that black holes
contain trapped surfaces, where both ingoing and outgoing light wavefronts are
converging, and that one can locate the surface of the black hole by {\em
marginal surfaces}, where outgoing light rays are instantaneously parallel. A
{\em trapping horizon} is a hypersurface foliated by marginal surfaces. Locally
classifying trapping horizons as future or past, and outer or inner, it was
proposed that a {\em future outer trapping horizon} characterizes
non-degenerate black holes. Some general results were that: there are future
trapped surfaces just inside such a horizon; the horizon is achronal, being
null only in the locally stationary case and otherwise spatial, assuming the
null energy condition; the marginal surfaces have spherical topology, assuming
the dominant energy condition; and the area $A$ of the horizon is
non-decreasing, $A'\ge0$, and increasing if spatial. The last property is
analogous to Hawking's area theorem, but for a practically locatable horizon.
Trapping horizons can be numerically located by so-called apparent-horizon
finders \cite{Schn,Tho}, which actually find marginal surfaces; they do not
check every surface in the hypersurface to see whether it is outer trapped, as
required by the definition of apparent horizon. Under smoothness assumptions
\cite{HE,KH}, apparent horizons are marginal surfaces, but not vice versa.
Incidentally, the resolution to the supposed black-hole-information paradox is
simple, using versions of the above results for matter violating the null
energy condition: as a black hole evaporates, the ingoing negative-energy
Hawking radiation causes the trapping horizon to shrink and become temporal, so
that information can cross it in both directions \cite{wh,evap}. There never
was a paradox, just a fundamental misunderstanding, that black holes are
usefully defined by event horizons.

In terms of trapping horizons, a comprehensive picture of black-hole dynamics
was first developed in spherical symmetry. In this case, there are local
definitions of active gravitational mass-energy $E$ \cite{sph} and surface
gravity $\kappa$ \cite{1st} which have many physically expected properties. The
gradient $dE$ of energy, expressed in terms of the energy-momentum of the
matter, divides naturally into an energy-supply term $A\psi$ and a work term
$wdV$ ($V=\frac43\pi R^3$, $A=4\pi R^2$) such that the energy flux $\psi$
vanishes on a trapping horizon if and only if it is null. This energy-balance
equation was called the unified first law for various reasons: projecting it
along the flow of a thermodynamic fluid yields a first law of relativistic
thermodynamics; projecting it along null infinity yields the Bondi energy-loss
equation, with $\psi$ reducing to the Bondi flux; and projecting $\psi$ along a
trapping horizon gives $\kappa A'/8\pi$, yielding an equation $E'=\kappa
A'/8\pi+wV'$ with the same form as the so-called first law for stationary black
holes, which is really analogous to the Gibbs equation rather than the first
law of thermodynamics. Including a zeroth law \cite{mg9}, the possible local
properties of dynamical black holes, independent of particular matter models,
are thereby known. This can also be achieved in cylindrical symmetry
\cite{cyl}, where the energy flux $\psi$ divides into contributions from the
matter and the gravitational radiation. A quasi-spherical approximation
\cite{qs,SH,gwbh,gwe} also allows generalizations of all these geometrical and
physical properties of black holes and gravitational radiation. An effective
energy tensor for the gravitational radiation can be given in all these cases.

Ashtekar \& Krishnan recently found an energy-balance law for {\em dynamical
horizons} \cite{AK1,AK2,Ash,AK3}, defined as spatial future trapping horizons.
This arose from earlier work on {\em isolated horizons}, types of null trapping
horizon for which generalizations of the classical laws of black-hole statics
were found \cite{ABF1,ABF2,AFK,ABD,ABL1,ABL2}. The new energy-balance law
describes how a general black hole grows due to the infalling matter and
gravitational radiation. In this article, these two threads are drawn together,
in particular deriving the first law for completely general trapping horizons,
so as to include both spatial (dynamical) and null (isolated) horizons, as well
as horizons of white holes, traversable wormholes, cosmological models and
evaporating black holes (\S VI). No displayed equation will assume any
restriction on the type of trapping horizon.

A terminological mismatch should be mentioned at the outset: the first law here
is what Ashtekar \& Krishnan called a balance equation for area or energy,
while their first law generalizes the so-called first law of black-hole
mechanics, involving new definitions of angular momentum and surface gravity,
which are not considered here. They gave an integral form using proper volume,
which becomes singular in the null limit, so here the first law is written in
differential form and in an alternative integral form, which are both regular
in the null limit (\S VII). An effective gravitational-radiation energy tensor
$\Theta$ and a preferred time vector $\chi$ are obtained, yielding
energy-tensor forms of the first law (\S VIII). A division into energy-supply
and work terms, generalizing the above structure in spherical or cylindrical
symmetry, allows a Gibbs-like equation involving a new definition of surface
gravity $\kappa$ (\S IX). The energy flux $\psi$ of the matter and
gravitational radiation satisfies a Clausius-like relation involving $\kappa$,
suggesting a definition of geometric entropy flux $2\pi\psi/\kappa$. Then it is
found that geometric entropy is conserved: the geometric entropy of the black
hole equals the geometric entropy supplied to the black hole by the infalling
matter and gravitational radiation (\S X). The results are preceded by brief
reviews of basic thermodynamics (\S II), the employed formalism of dual-null
dynamics (\S III), the definition of trapping horizon (\S IV) and the area and
signature laws (\S V), and followed by a Conclusion (\S XI). See \cite{bhd2}
for a short version and \cite{BF,Yoo} for different but related approaches.
Standard Einstein gravity is assumed, though the ideas generalize.

\section{Basic thermodynamics}

It seems appropriate to begin with a brief summary of basic thermodynamics, due
to the parallels often drawn for black holes, and the fact that they are not
always accurate. See fuller treatments \cite{Eck,MR,dGM,th} and beware any
source which formulates the laws of thermodynamics using state-space
differentials $d$ and meaningless derivatives $\delta$.

In classical thermodynamics, the basic quantities are temperature $\vartheta$,
heat supply $Q$, work $W$, internal energy (actually thermal energy or simply
heat) $H$ and entropy $S$. The classical first law is
\begin{equation}
\dot H=\dot Q+\dot W\label{first0}
\end{equation}
where the dot denotes the material or comoving derivative. For instance, for an
inviscid fluid, the work is given by $\dot W=-p\dot V$, where $V$ is the volume
and $p$ the pressure of the fluid, so that the first law reads
\begin{equation}
\dot H=\dot Q-p\dot V.
\end{equation}
The classical second law, originally due to Clausius, who used it to define
entropy, is
\begin{equation}
\dot S\ge\dot Q/\vartheta.
\end{equation}
These integral forms of the laws respectively require the pressure and
temperature to be spatially constant.

The entropy may be divided into entropy supply $S_\circ$, given by
\begin{equation}
\dot S_\circ=\dot Q/\vartheta
\end{equation}
and entropy production $S-S_\circ$. Then the second law may be written as
\begin{equation}
\dot S\ge\dot S_\circ\label{second0}
\end{equation}
which expresses entropy production. In words: $S$ is the entropy of the system,
where system means a comoving volume of material, and $S_\circ$ is the entropy
supplied to the system. Thus the second law implies that the total entropy of
an isolated system, such as the whole universe, cannot decrease. Here it should
be stressed that dynamical black holes are not isolated systems, since they
absorb energy and entropy. Then the property that black holes have
non-decreasing area, $A'\ge0$, normally called the second law of black-hole
mechanics, is actually not analogous to the second law of thermodynamics.
Entropy production and entropy increase have entirely different meanings for
non-isolated systems.

Equality in the second law holds in thermostatics, traditionally called
equilibrium thermodynamics or reversible thermodynamics. In the thermostatic
case, the first and second laws for an inviscid fluid imply
\begin{equation}
\dot H=\vartheta\dot S-p\dot V\label{gibbs0}
\end{equation}
which is the Gibbs equation, or rather its material or comoving form. Note that
what is normally called the first law of black-hole mechanics for stationary
black holes \cite{BCH}, involving area $A=4S$ and surface gravity
$\kappa=2\pi\vartheta$, is actually analogous to the Gibbs equation, rather
than the first law of thermodynamics. The latter does not involve temperature
or entropy, but simply expresses energy balance.

Thermodynamics can be formulated as a local field theory, with $H$ and $S$
replaced by thermal energy density and entropy density respectively, $Q$
replaced by a thermal flux vector $q$ such that
\begin{equation}
\dot Q=-\oint{*}n\cdot q\label{heat}
\end{equation}
and $S_\circ$ replaced by an entropy flux vector $\varphi=q/\vartheta$ such
that
\begin{equation}
\dot S_\circ=-\oint{*}n\cdot\varphi\label{entropy0}
\end{equation}
where the integrals are over a surface bounding the system, with vector area
element ${*}n$. In terms of these and other fields, e.g.\ density, velocity and
stress for a fluid, the first and second laws and the Gibbs equation can be
localized \cite{Eck,MR,dGM,th}. These three localized equations or inequalities
can then be used to derive dissipative relations, in the simplest case the
Fourier equation for $q$ and the Newtonian-fluid equation for the viscous
stress, leading to the Navier-Stokes equation. Thus it should be stressed that
the first law and comoving Gibbs equation are still assumed fundamentally and
fruitfully in true (non-equilibrium) thermodynamics as well as in
thermostatics. Widespread folklore to the contrary is sometimes used to argue
that the so-called first law of black-hole mechanics, obtained as a property of
stationary black holes, should not be expected to generalize to dynamical black
holes. Again, this does not constitute a correct analogy with true
thermodynamics.

\section{Dual-null dynamics}

Trapping horizons are generally defined as hypersurfaces which may have any
causal nature, foliated by marginal surfaces. To study them, it is useful to
employ the formalism of dual-null dynamics \cite{dn,dne}, describing two
families of null hypersurfaces, intersecting in a two-parameter family of
transverse spatial surfaces, as summarized in this section. There are various
reasons: marginal surfaces are defined as extremal surfaces of null
hypersurfaces; a spatial trapping horizon locally determines a unique dual-null
foliation, generated from the marginal surfaces in the null normal directions;
and the null limit, where a dynamical horizon reduces to an isolated horizon,
is naturally included in the formalism, whereas more conventional treatments of
spatial hypersurfaces become degenerate in the null limit, basically because
normal vectors become tangent. For a null trapping horizon, the dual-null
foliation is not unique, so subtleties remain in describing partial spatial,
partially null trapping horizons.

Denoting the space-time metric by $g$ and labelling the null hypersurfaces by
coordinates $x^\pm$ which increase to the future, the normal 1-forms
\begin{equation}
n^\pm=-dx^\pm
\end{equation}
therefore satisfy
\begin{equation}
g^{-1}(n^\pm,n^\pm)=0.
\end{equation}
The relative normalization of the null normals
may be encoded in a function $f$ defined by
\begin{equation}
e^f=-g^{-1}(n^+,n^-)
\end{equation}
where the metric sign convention is that spatial metrics are positive definite.
Some readers may prefer to write $g^{+-}=g^{-1}(n^+,n^-)$ for more manifest
invariance and remember that $g^{+-}<0$. The induced metric on the transverse
surfaces, the spatial surfaces of intersection, is found to be
\begin{equation}
h=g+2e^{-f}n^+\otimes n^-
\end{equation}
where $\otimes$ denotes the symmetric tensor product. The dynamics are
generated by two commuting evolution vectors $u_\pm$:
\begin{equation}
[u_+,u_-]=0
\end{equation}
where the brackets denote the Lie bracket or commutator. Thus there is an
integrable evolution space spanned by $(u_+,u_-)$. There are two shift vectors
\begin{equation}
s_\pm=\bot u_\pm
\end{equation}
where $\bot$ indicates projection by $h$. The null normal vectors
\begin{equation}
l_\pm=u_\pm-s_\pm=e^{-f}g^{-1}(n^\mp)
\end{equation}
are future-null and satisfy
\begin{eqnarray}
g(l_\pm,l_\pm)&=&0\\
g(l_+,l_-)&=&-e^{-f}\\
l_\pm\cdot dx^\pm&=&1\\
l_\pm\cdot dx^\mp&=&0\\
\bot l_\pm&=&0
\end{eqnarray}
where a dot denotes symmetric contraction. In a coordinate basis
$(u_+,u_-,e_a)$ such that $u_\pm=\partial/\partial x^\pm$, where
$e_a=\partial/\partial x^a$ is a basis for the transverse surfaces, the metric
takes the form
\begin{equation}
g=h_{ab}(dx^a+s_+^adx^++s_-^adx^-)\otimes(dx^b+s_+^bdx^++s_-^bdx^-)
-2e^{-f}dx^+\otimes dx^-.
\end{equation}
Then $(h,f,s_\pm)$ are configuration fields and the independent momentum fields
are found to be linear combinations of the following transverse tensors:
\begin{eqnarray}
\theta_\pm&=&{*}L_\pm{*}1\\ \sigma_\pm&=&\bot L_\pm h-\theta_\pm h\\
\nu_\pm&=&L_\pm f\\ \omega&=&\textstyle{1\over2}e^fh([l_-,l_+])
\end{eqnarray}
where ${*}$ is the Hodge operator of $h$ and $L_\pm$ is shorthand for the Lie
derivative along $l_\pm$. Then the functions $\theta_\pm$ are the expansions,
the traceless bilinear forms $\sigma_\pm$ are the shears, the 1-form $\omega$
is the twist, measuring the lack of integrability of the normal space, and the
functions $\nu_\pm$ are the inaffinities, measuring the failure of the null
normals to be affine. The fields $(\theta_\pm,\sigma_\pm,\nu_\pm,\omega)$
encode the extrinsic curvature of the dual-null foliation. These extrinsic
fields are unique up to duality $\pm\mapsto\mp$ and diffeomorphisms
$x^\pm\mapsto\tilde x^\pm(x^\pm)$ which relabel the null hypersurfaces. It will
also be convenient to use the normal fundamental forms
\begin{equation}
\zeta_{(\pm)}=e^{-f}\bot((n^\mp\cdot\nabla^\sharp)n^\pm)=Df/2\mp\omega
\end{equation}
where $\nabla$ and $D$ are the covariant derivatives of $g$ and $h$
respectively, and $\nabla\wedge n^\pm=0$ has been used. Here $(\pm)$ indicates
a label, not an index; $\zeta_{(\pm)}$ are two generally distinct transverse
1-forms. One can compose them as a 2-form $\beta$ in the normal space, defined
by $\beta(\mu,\nu)=\bot((\mu^\sharp\cdot\nabla)\nu)$ for normal 1-forms
$(\mu,\nu)$, with components $\beta^{\pm\mp}=e^f\zeta_{(\mp)}$,
$\beta^{\pm\pm}=0$, but such notation becomes cumbersome. Likewise, one can
compose the expansions and shears into a second fundamental form, but it is
more convenient to separate them.

\begin{figure}
\includegraphics[width=12cm]{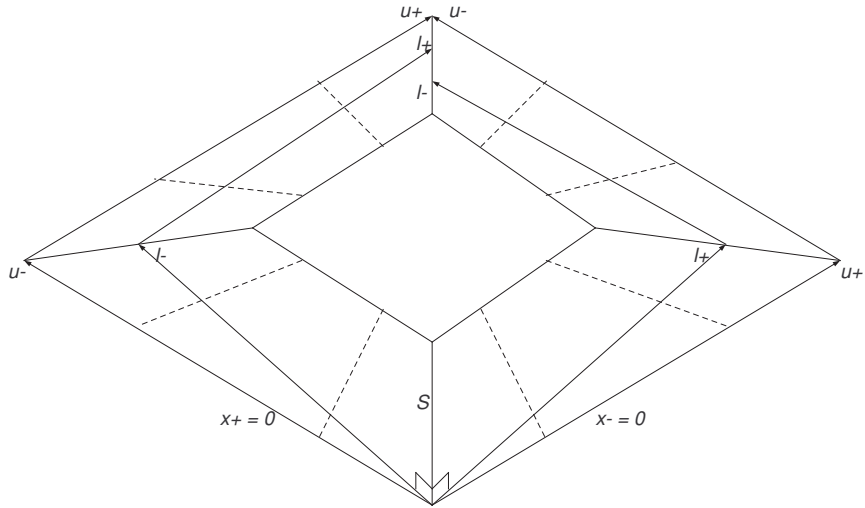}
\caption{A dual-null foliation: the commuting evolution vectors
$u_\pm=\partial/\partial x^\pm$ generate the transverse surfaces $S$, while
their null normal projections $l_\pm$ generally do not commute.} \label{dnf}
\end{figure}

One subtle point concerns the evolution vectors $u_\pm$ versus the the null
normal vectors $l_\pm$, differing by the shift vectors $s_\pm$. In a numerical
evolution, one would be evolving using the field equations with $\bot
L_{u_\pm}$ derivatives on the left-hand side, since such Lie propagation of a
point takes it to other points with the same angular coordinates. In
particular, evolving along $u_+$ then $u_-$ takes one to the same point as
evolving along $u_-$ then $u_+$, since they commute, as depicted in
Fig.~\ref{dnf}. However, $l_\pm$ generally do not commute (as measured by
$\omega$), so that evolving along $l_+$ then $l_-$ takes one to a generally
different point, though in the same transverse surface, as evolving along $l_-$
then $l_+$. On the other hand, for analytical purposes it is easier to use
$l_\pm$ than $u_\pm$, writing the field equations with $\bot L_\pm$ derivatives
on the left-hand side. The same issue exists in the 3+1 formalism, where the
evolution vector differs by the lapse function and shift vector from the unit
normal vector. Another subtle point is that $l_\pm$ are not general tetrad
vectors, since they are defined in terms of $n^\mp$, which must be closed. In
particular, this means that $f$ cannot be fixed to zero for a general dual-null
foliation. Another way to see this is that its derivatives $L_\pm L_\mp f$ are
determined by the Einstein system in terms of the free initial data \cite{dne},
even in spherical symmetry.

The dual-null Hamilton equations and integrability conditions for vacuum
Einstein gravity were derived previously \cite{dne}, with matter terms added
subsequently \cite{gwbh}. Denoting projections of the energy tensor $T$ by
$T_{\pm\pm}=T(l_\pm,l_\pm)$ and $T_{+-}=T(l_+,l_-)$, the relevant components of
the field equations are just
\begin{eqnarray}
L_\pm\theta_\pm&=&-\nu_\pm\theta_\pm-\theta_\pm^2/2 -||\sigma_\pm||^2/4 -8\pi
T_{\pm\pm}\label{focus1}\\
L_\mp\theta_\pm&=&-\theta_+\theta_- -e^{-f}\left(\Re/2-|\zeta_{(\pm)}|^2
+D\cdot\zeta_{(\pm)}^\sharp\right)+8\pi T_{+-}\label{focus2}
\end{eqnarray}
where $\Re$ is the Ricci scalar of $h$ (conventionally positive for spheres), a
sharp ($\sharp$) denotes the contravariant dual with respect to
$h^{-1}=h^\sharp$ (index raising), $|\zeta|^2=\zeta\cdot\zeta^\sharp$ and
$||\sigma||^2=\sigma:\sigma^\sharp$, where the colon denotes double symmetric
contraction. Units are such that Newton's gravitational constant is unity. The
first equation is the well-known null focusing equation and the second has been
called the cross-focusing equation \cite{bhd,bhs}. The null energy condition
implies
\begin{equation}
T_{\pm\pm}\ge0\label{null}
\end{equation}
and the dominant energy condition additionally implies
\begin{equation}
T_{+-}\ge0.\label{dominant}
\end{equation}

\section{Trapping horizons}

The dual-null formalism may be applied to one-parameter families of transverse
surfaces generated by a vector
\begin{equation}
\xi=\xi^+l_++\xi^-l_-\qquad D\xi^\pm=0.
\end{equation}
This means that $\xi=\partial/\partial x$ is normal to the constant-$x$
transverse surfaces, so that $\xi^\pm$ can be taken outside transverse surface
integrals $\oint$. The area of the transverse surfaces is
\begin{equation}
A=\oint{*}1
\end{equation}
and the area radius
\begin{equation}
R=\sqrt{A/4\pi}
\end{equation}
is often more convenient. The Hawking energy \cite{Haw}
\begin{equation}
E=\frac{R}{16\pi}\oint{*}\left(\Re+e^f\theta_+\theta_-\right)\label{energy}
\end{equation}
will be used as a measure of the active gravitational mass-energy on a
transverse surface. On a stationary black-hole horizon, it is also known as the
irreducible mass: the mass which must remain even if rotational or electrical
energy is extracted.

Consider a trapping horizon generated by a vector $\xi=\partial/\partial x$, so
that the constant-$x$ surfaces are marginal surfaces, where one of the null
expansions $\theta_\pm$ vanishes. This leaves the freedom to relabel the
marginal surfaces, $x\mapsto\hat x(x)$, under which all the key equations will
be manifestly invariant. Equations holding on a trapping horizon will be
denoted by the weak equality symbol $\cong$. Initially, the case
$\theta_+\cong0$ will be considered in detail, with the case $\theta_-\cong0$
included subsequently. The fundamental equation describing the evolution of a
trapping horizon is
\begin{equation}
0\cong L_\xi\theta_+=\xi^+L_+\theta_++\xi^-L_-\theta_+\label{horizon1}
\end{equation}
where $L_\xi$ denotes the Lie derivative along $\xi$. This will be used
together with the Einstein equation to derive the first law for any trapping
horizon. Even without the Einstein equation, one can use it to deduce
relationships between the signs of $\xi^\pm$ and $L_\pm\theta_+$, which
determine the causal nature of the trapping horizon and whether its area
increases or decreases, via
\begin{equation}
L_\xi A=\oint{*}(\xi^+\theta_++\xi^-\theta_-).\label{area}
\end{equation}
For clarity, all such inequalities are collected in the next section, so that
the remainder of the article applies to any trapping horizon.

\section{Area and signature laws}

Trapping horizons were previously classified \cite{bhd} into one of four
non-degenerate types: future (respectively past) if $\theta_-<0$ (respectively
$\theta_->0$) and outer (respectively inner) if $L_-\theta_+<0$ (respectively
$L_-\theta_+>0$) on the trapping horizon $\theta_+\cong0$. For each type, there
are trapped surfaces ($\theta_+\theta_->0$) to one side of the horizon and
untrapped surfaces ($\theta_+\theta_-<0$) to the other side, which is not
guaranteed if the inequalities are relaxed even to non-strict inequalities. The
causal type of the horizon is determined pointwise by the relative signs of
$\xi^\pm$: spatial if they have opposite signs, null if one vanishes and the
other does not, and temporal if they have the same (non-zero) sign. Since the
null energy condition (\ref{null}) and focusing equation (\ref{focus1}) imply
$L_+\theta_+\le0$ on the trapping horizon, it follows from the fundamental
equation (\ref{horizon1}) that outer trapping horizons are achronal (spatial or
null), while inner trapping horizons are causal (temporal or null); the
signature law \cite{bhd}. Furthermore, fixing the orientation of $\xi$ by
$\xi^+>0$, it follows from (\ref{area}) that the area of a future outer or past
inner trapping horizon is non-decreasing, $L_\xi A\ge0$, while the area of a
past outer or future inner trapping horizon is non-increasing, $L_\xi A\le0$;
the area law \cite{bhd}. As corollaries, the horizon is null and has
instantaneously constant area if and only if the ingoing energy density
$T_{++}+\Theta_{++}$ vanishes, where $\Theta_{++}=||\sigma_+||^2/32\pi$
(\ref{Theta0}) can be understood subsequently as the effective energy density
of ingoing gravitational radiation.

As mentioned in the Introduction, non-degenerate black holes may be
characterized by future outer trapping horizons. Ashtekar \& Krishnan instead
defined dynamical horizons as spatial future trapping horizons. Then
$\theta_-<0$ on the horizon and $\xi^\pm$ have opposite signs. Choosing the
orientation of $l_\pm$ such that $l_+$ is outward and $l_-$ inward, $\xi^+>0$,
$\xi^-<0$ and it follows directly from (\ref{area}) that the area is
increasing, $L_\xi A>0$, a strict version of the above area law. Actually, for
black holes, one is normally interested in future trapping horizons which are
either spatial (dynamical) or future-null (isolated), or partially spatial and
partially null. In such cases $\xi^+>0$ and $\xi^-\le0$, which immediately
gives the non-strict area law $L_\xi A\ge0$. Note also from the fundamental
equation (\ref{horizon1}) that a dynamical horizon satisfies $L_-\theta_+\le0$
under the null energy condition, so it is either a future outer trapping
horizon or degenerate. The degenerate cases allow dynamical horizons in
space-times without trapped surfaces \cite{Sen}, reflecting the need for
something like the outer condition to characterize a black hole. In practice,
the outer horizon of a black hole is likely to satisfy both definitions, except
when it becomes stationary or instantaneously stationary.

An evaporating black hole may also be described using trapping horizons. The
only difference with the above discussion is that the null energy condition is
violated by Hawking radiation, for which the ingoing radiation has negative
energy density, $T_{++}<0$. Assuming that this dominates the positive energy
density of ingoing gravitational radiation, $T_{++}+\Theta_{++}<0$, the
focusing equation (\ref{focus1}) implies $L_+\theta_+>0$ on the trapping
horizon. For an outer horizon, the fundamental equation (\ref{horizon1})
implies that $\xi^\pm$ have the same sign, so that the horizon is temporal,
while (\ref{area}) shows that the area is decreasing for a future horizon,
$L_\xi A<0$. Thus the black-hole horizon is shrinking and two-way traversable.
Clearly matter can escape from an evaporating black hole. The strange belief
that information cannot escape from an evaporating black hole seems to be based
on the impractical event-horizon definition of black hole as a region of no
escape.

\section{First law: energy flux and work}

Henceforth completely general trapping horizons will be considered, so that all
the following displayed equations will apply not only to outer black-hole
horizons under the usual energy conditions, but to inner black-hole horizons,
white holes, cosmological horizons, wormhole mouths and evaporating black
holes. Expanding the fundamental relation (\ref{horizon1}) using the focusing
equations (\ref{focus1}--\ref{focus2}) yields
\begin{equation}
0\cong L_\xi\theta_+\cong-\xi^+\left(8\pi
T_{++}+||\sigma_+||^2/4\right)+\xi^-\left(8\pi
T_{+-}-e^{-f}\left(\Re/2-|\zeta|^2+D\cdot\zeta^\sharp\right)\right)\label{horizon2}
\end{equation}
where $\zeta=\zeta_{(+)}$ temporarily simplifies the notation. Multiplying by
$e^f/8\pi$ and integrating over the transverse surfaces, using the Gauss-Bonnet
theorem $\oint{*}\Re=8\pi$, the Gauss divergence theorem
$\oint{*}D\cdot\alpha=0$ and rearranging yields
\begin{equation}
\xi^+\oint{*}e^f\left(T_{++}+\frac{||\sigma_+||^2}{32\pi}\right)
-\xi^-\oint{*}\left(e^fT_{+-}+\frac{|\zeta|^2}{8\pi}\right)\cong-\xi^-/2.
\label{horizon3}
\end{equation}
Since $\xi^-=0$ in the null case, this shows that both $T_{++}$ and $\sigma_+$
must vanish on a null horizon \cite{bhd}, assuming the null energy condition.
Here spherical topology has been assumed; otherwise, for compact orientable
transverse surfaces, the right-hand side of (\ref{horizon3}) is multiplied by
$1-\gamma$, where $\gamma$ is the genus or number of handles. Then the dominant
energy condition implies $\gamma\le1$, leading to the topology law \cite{bhd}:
the transverse surfaces are either spherical or toroidal, the latter case
requiring very special conditions (including vanishing Gaussian curvature) and
anyway being excluded for (non-degenerate) outer trapping horizons,
$L_-\theta_+<0$.

The Hawking mass-energy (\ref{energy}) satisfies
\begin{equation}
E\cong R/2
\end{equation}
on a trapping horizon, which can be regarded as a generalization of irreducible
mass from stationary to non-stationary black holes, since the area law ensures
its irreducibility under the null energy condition. Then the identity
(\ref{horizon3}) yields
\begin{equation}
L_\xi E\cong\frac{L_\xi R}2\cong\left[
{}-\frac{\xi^+}{\xi^-}\oint{*}e^f\left(T_{++}+\frac{||\sigma_+||^2}{32\pi}\right)
+\oint{*}\left(e^fT_{+-}+\frac{|\zeta|^2}{8\pi}\right)\right]L_\xi R.
\label{first1}
\end{equation}
This is a dual-null differential version of the energy-balance law found by
Ashtekar \& Krishnan \cite{AK1,AK2}, as compared more explicitly in the next
section. One may fix the normalization $f\cong0$ and, for a spatial horizon,
one may fix the scaling of the null normals such that $\xi^+/\xi^-\cong-1$ and
the generating vector $\xi$ such that $L_\xi R\cong1$. In the following, all
gauge freedom will be retained for generality, but readers may wish on a first
reading to mentally set $f\cong0$ and, if interested only in spatial trapping
horizons, $\xi^+/\xi^-\cong-1$. The four terms are all geometrical invariants
of the dual-null foliation, as shown explicitly below, and therefore of the
horizon (as an embedded hypersurface) unless it becomes null. Since the
formalism is manifestly covariant on the transverse surfaces, checking
invariance reduces to writing $e^f=-g^{+-}$ and matching $\pm$ indices.

The four terms in parentheses in(\ref{first1}) are all manifestly positive,
assuming the dominant energy condition. The $T_{++}$ term gives the energy flux
of the matter propagating in the null direction into the horizon. Consequently
it is natural to interpret the $\sigma_+$ shear term as giving the energy flux
of the transverse gravitational radiation propagating in the null direction
into the horizon. This term has the same form as that of the Bondi flux of
gravitational radiation at null infinity \cite{mon,inf}, the same form as a
localized energy flux of gravitational radiation in a quasi-spherical
approximation \cite{gwbh}, and a similar form to the energy flux of linearized
gravitational radiation in the high-frequency approximation \cite{MTW}, so its
physical interpretation seems sound. The $T_{+-}$ term gives a matter energy
density, so the $\zeta$ term can be interpreted as giving a corresponding
gravitational energy density. Ashtekar \& Krishnan interpreted it as also due
to gravitational radiation, and here it will be interpreted as an energy
density of longitudinal gravitational radiation. This is much less familiar
than transverse gravitational radiation and is absent in all the above
approximations, but the interpretation can be understood in a spin-coefficient
formulation, to be presented elsewhere \cite{bhd4}. It should be mentioned that
there is a widespread belief that longitudinal gravitational radiation does not
exist in Einstein gravity, apparently due to an argument in linearized theory
that the longitudinal modes are purely gauge-dependent, and the fact that only
the transverse mode contributes to the Bondi flux. On the first point, Szekeres
characterized ingoing and outgoing, transverse and longitudinal gravitational
radiation by their effect via the geodesic equation on a ``gravitational
compass'' of test particles \cite{Sze,Nol}. On the second point, it can be
shown that the energy densities of the outgoing transverse and longitudinal
modes fall off near future null infinity as $1/R^2$ and $1/R^4$ respectively
\cite{inf,bhd4}.

The expression (\ref{first1}) separates the first term in parentheses, which
vanishes for null horizons (assuming the null energy condition), from the
second term in parentheses, which is generally non-zero for horizons of any
causal nature. This separation need not appear for spatial trapping horizons,
but it will be stressed in the following, since the null case is a physically
important limit, where dynamical horizons reduce to isolated horizons, or more
prosaically, where a growing black hole ceases to grow.

The next task is to write the new law in a more manifestly invariant form. The
spherically symmetric case is a useful guide; there the unified first law was
found as $dE=A\psi+wAdR$ for certain invariants $\psi$ and $w$ of the matter
energy tensor \cite{1st}. The corresponding formulae read
\begin{eqnarray}
w_m&=&-\hbox{trace}\,T/2\\
\psi_m&=&T\cdot (dR)^\sharp+w_mdR
\end{eqnarray}
where the trace is in the normal space and the subscript $m$ is introduced to
refer to the matter. In the current generalized context, these quantities are
invariants of the dual-null foliation. Explicitly,
\begin{eqnarray}
w_m&=&e^fT_{+-}\\
(\psi_m)_\pm&=&-e^fT_{\pm\pm}L_\mp R
\end{eqnarray}
where $\psi=\psi_+dx^++\psi_-dx^-$. Comparing with the first law
(\ref{first1}), one can define corresponding quantities for the gravitational
radiation by
\begin{eqnarray}
w_g&=&\frac{|\zeta|^2}{8\pi}\\
(\psi_g)_\pm&=&-\frac{e^f||\sigma_\pm||^2L_\mp R}{32\pi}.
\end{eqnarray}
As in the spherically symmetric case \cite{1st}, one may call
\begin{equation}
w=w_m+w_g
\end{equation}
the {\em work density} and
\begin{equation}
\psi=\psi_m+\psi_g\label{flux1}
\end{equation}
the {\em energy flux} 1-form.

On a trapping horizon with $\theta_+\cong0$, it follows that $L_+R\cong0$,
$\psi_-\cong0$, $\xi\cdot\psi\cong\xi^+\psi_+$ and $L_\xi R\cong\xi^-L_-R$,
yielding
\begin{equation}
\xi\cdot\psi\cong{}-\frac{\xi^+}{\xi^-}
e^f\left(T_{++}+\frac{||\sigma_+||^2}{32\pi}\right)L_\xi R.\label{flux2}
\end{equation}
Thus the first law (\ref{first1}) becomes
\begin{equation}
L_\xi E\cong\oint{*}\xi\cdot\psi+\oint{*}wL_\xi R.\label{first2}
\end{equation}
This is the {\em first law of black-hole dynamics}, in the desired
geometrically invariant form. (The Lie derivative $L_\xi$ acting on integral
scalars like $E$ and $R$ is just the partial derivative $\partial/\partial x$).
The energy flux $\xi\cdot\psi$ vanishes if the horizon is null, while the work
density $w$ is generally non-zero for horizons of any causal nature. The two
terms $\oint{*}\xi\cdot\psi$ and $\oint{*}wL_\xi R$ may be called respectively
the {\em energy-supply} and {\em work} terms, by analogy with the first law of
thermodynamics (\ref{first0}).

The above derivation of the first law applied to a trapping horizon with
$\theta_+\cong0$. For a trapping horizon with $\theta_-\cong0$, one obtains the
same formula with a different $w$. For completeness one can define
\begin{equation}
w_{(\pm)}=e^fT_{+-}+\frac{|\zeta_{(\pm)}|^2}{8\pi}
\end{equation}
and use $w\cong w_{(\pm)}$ for horizons with $\theta_\pm\cong0$. Recall that
one can anyway fix the normalization $f\cong0$, in which case
$\omega\cong\zeta_{(-)}\cong-\zeta_{(+)}$. On the other hand, the same $\psi$
appears in both cases, indicating that both components $\psi_\pm$ are correct,
i.e.\ that the energy densities given by the first bracket in (\ref{first1}),
which at first sight are just scalars, can be naturally derived from an energy
flux 1-form $\psi$ as $\xi\cdot\psi$.

\section{Integral forms of the first law}

Ashtekar \& Krishnan derived an integral form of the first law, using proper
volume, whereas the first law (\ref{first2}) derived above is in differential
form. For comparison, it can be written simply as
\begin{equation}
L_\xi E\cong\oint{*}\epsilon L_\xi R
\end{equation}
where
\begin{equation}
\epsilon=\frac{\xi\cdot\psi}{L_\xi R}+w
\end{equation}
is the combined {\em energy density}, where the division into energy-supply and
work terms can be ignored in this section. It can be independently divided into
matter and gravitational-radiation terms, $\epsilon=\epsilon_m+\epsilon_g$, in
the obvious way, yielding the explicit expressions
\begin{eqnarray}
\epsilon_m&\cong&e^f\left(T_{+-}-\frac{\xi^\pm}{\xi^\mp}T_{\pm\pm}\right)
\label{ed1}\\
\epsilon_g&\cong&\frac1{32\pi}
\left(4|\zeta_{(\pm)}|^2-e^f\frac{\xi^\pm}{\xi^\mp}||\sigma_\pm||^2\right)
\label{ed2}
\end{eqnarray}
for a trapping horizon with $\theta_\pm\cong0$.

A corresponding integral form of the first law is
\begin{equation}
[E]\cong\int{*}\epsilon L_\xi R\wedge dx
\end{equation}
which expresses the change in $E$ along the horizon, from one marginal surface
to another. In this article, $\int$ always denotes such a hypersurface integral
between transverse surfaces, $\oint$ always denotes a transverse surface
integral and $[\,]$ denotes the change in such a quantity between transverse
surfaces. This manifestly invariant expression uses the generator-volume
element ${*}1\wedge dx$. Alternatively, if one wishes to use the proper-volume
element
\begin{equation}
\tilde{*}1={*}1\wedge ds={*}\sqrt{g_{xx}}\wedge dx
\end{equation}
where $s$ is arc length along the horizon-generating vector
$\xi=\partial/\partial x$ and
\begin{equation}
\sqrt{g_{xx}}=\sqrt{g(\xi,\xi)}=\partial s/\partial x
\end{equation}
is the corresponding scale factor, then the integral first law can be written
as
\begin{equation}
[E]\cong\int\tilde{*}\tilde\epsilon
\end{equation}
where
\begin{equation}
\tilde\epsilon=\epsilon L_\eta R=\eta\cdot\psi+wL_\eta R
\end{equation}
is the {\em proper energy density} and
\begin{equation}
\eta=\xi/\sqrt{g_{xx}}=\partial/\partial s
\end{equation}
is the vector which is parallel to a normal generating vector and
differentiates with respect to arc length, $\eta\cdot ds=1$. Note that
$\tilde{*}1$, $s$ and $\eta$ are all independent of relabellings of the
marginal surfaces, $x\mapsto\hat x(x)$.

The last form is a compact version of the Ashtekar-Krishnan energy-balance law,
so one can finally check consistency. They assumed gauge choices which here
correspond to $f\cong0$, $\xi^+/\xi^-\cong-1$ and $L_\xi R\cong1$, mentioned
after (\ref{first1}) as admissible for a spatial horizon, and their null
normals are $\sqrt2l_\pm$, but it is now straightforward to check their
expressions against the explicit expressions (\ref{ed1}--\ref{ed2}), which
reduce to $\epsilon_m\cong T_{++}+T_{+-}$ and
$\epsilon_g\cong(||\sigma_+||^2+4|\zeta_{(+)}|^2)/32\pi$ for $\theta_+\cong0$,
with $\tilde\epsilon$ reducing to $\epsilon/\sqrt{g_{xx}}$. They also gave a
more general form for $[x/2]$, not fixing $L_\xi R$, discussing $x=R$ as a
special case; this can easily be reproduced here simply by dividing both sides
of the first law (in whatever form) by $L_\xi R$. Note some notational
pitfalls: their lapse $N$ is $1/\sqrt{g_{xx}}$ here, their coordinate $r$ is
$x$ here, their $\sigma$ is $\sigma_+/\sqrt2$ here, and their permissible
vectors $\xi$ would here be $l_+\sqrt{2/g_{xx}}$, rescaling the outward null
normal $l_+$ rather than the horizon-generating vector $\xi$ (to
$\eta=\xi/\sqrt{g_{xx}}$). Either way, the rescaled vector is ill-defined if
the trapping horizon becomes null, $g_{xx}\to0$. In this limit, the
proper-volume element vanishes, while the apparently ill-defined proper energy
density $\tilde\epsilon$ turns out to be finite. This suggests using either the
generator-volume or differential form to deal with partially spatial, partially
null trapping horizons.

\section{Energy-tensor forms: effective gravitational-radiation energy tensor}

Dividing the integrated energy fluxes into those due to the matter and the
gravitational radiation in the obvious way,
\begin{equation}
[E]=[E]_m+[E]_g.
\end{equation}
Ashtekar \& Krishnan stressed that one could obtain the integrated matter flux
as an equation which here would be
$[E]_m\cong\int\tilde{*}T(l_+\sqrt{2/g_{xx}},\hat\tau)$ in their gauge choice,
where $\hat\tau$ is the unit normal vector to a spatial trapping horizon.
However, unit $\hat\tau$ does not exist for null trapping horizons.
Nevertheless a natural normal does exist for any trapping horizon, namely the
vector $\tau$ dual to the generating vector $\xi$ in the normal space:
\begin{equation}
\tau=\hat{*}\xi
\end{equation}
where $\hat{*}$ is the vectorial Hodge operator of the normal space, with
orientation chosen so that $\tau=\xi^+l_+-\xi^-l_-$, meaning that $\tau$ is
future-pointing for outward-pointing $\xi$. Then $\tau$ is normal to the
horizon, $g(\xi,\tau)=0$, $\bot\tau=0$, has equal and opposite normalization
$g(\tau,\tau)=-g(\xi,\xi)$ and is regular in the null limit, becoming null
itself, $\tau\to\xi$.

\begin{figure}
\includegraphics[width=18cm]{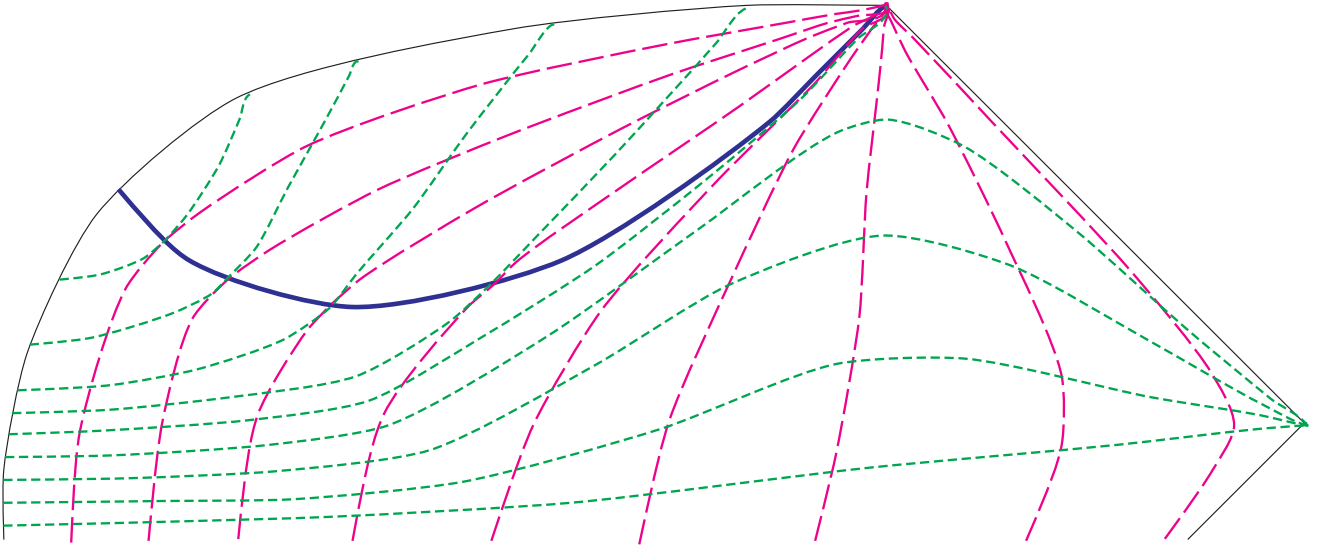}
\caption{(colour online). Gravitational collapse: Penrose diagram of typical
black-hole formation satisfying cosmic censorship, indicating the trapping
horizon (blue bold line), null infinity (straight lines) and the centre $R=0$
(curved line), which is regular outside the trapping horizon and singular and
spatial inside. Advanced time $x^+$ runs diagonally up-rightwards and retarded
time $x^-$ runs diagonally up-leftwards. Flow lines of the area-radius vector
$(dR)^\sharp$ (green short dashes) and the Killing-like vector $\chi$ (magenta
long dashes) are also indicated. Note the causal switch-over between $\chi$ and
$(dR)^\sharp$ at the trapping horizon.} \label{collapse}
\end{figure}

In spherical symmetry, the Kodama vector $\chi$ provides a preferred flow of
time, reducing to the stationary Killing vector for Schwarzschild and
Reissner-Nordstr\"om black holes. It has a dual relation to the energy $E$
which can be written as $L_\xi E=AT(\chi,\tau)$, for any normal vector $\xi$
and its orthogonal dual $\tau$. This vector can be generalized by
\begin{equation}
\chi=\hat{*}(dR)^\sharp\label{chi}
\end{equation}
or the curl of $R$ in the normal space, with components
$\chi=e^f(L_+R\,l_--L_-R\,l_+)$. Then $\chi$ is orthogonal to $R$ and the
transverse surfaces, $\chi\cdot dR=0$, $\bot\chi=0$, has normalization
$g(\chi,\chi)=-g^{-1}(dR,dR)$ and becomes null on a trapping horizon,
$g(\chi,\chi)\cong0$, with $\chi\cong\pm(dR)^\sharp$ for $\theta_\pm\cong0$.
Flow lines of $\chi$ and $(dR)^\sharp$ are sketched in Fig.~\ref{collapse} for
typical gravitational collapse to a black hole, assuming cosmic censorship; for
a comprehensively analyzed case, see Christodoulou \cite{Chr} for the massless
Klein-Gordon field in spherical symmetry. For a future outer trapping horizon
\cite{bhd}, the area-radius vector $(dR)^\sharp$ is spatial and $\chi$ is
temporal just outside the horizon, and $(dR)^\sharp$ is temporal and $\chi$
spatial just inside the horizon.

In terms of these vectors, there is a remarkably simple and manifestly
invariant expression for the matter energy density:
\begin{equation}
\epsilon_mL_\xi R=T(\chi,\tau).
\end{equation}
This holds for any foliation of spatial surfaces in any space-time, generated
by a normal vector $\xi$ with orthogonal dual $\tau$, without any gauge
conditions. Then the integrated matter flux is
\begin{equation}
[E]_m\cong\int{*}T(\chi,\tau)\wedge dx.
\end{equation}
Here the rescaling freedom $x\mapsto\hat x(x)$ in $\xi$ and therefore $\tau$ is
cancelled by $dx$ to leave an invariant expression. With the Ashtekar-Krishnan
gauge conditions $f\cong0$, $\xi^+/\xi^-\cong-1$ and $L_\xi R\cong1$, one finds
that $\chi$ reduces to $l_+\sqrt{2/g_{xx}}$ on a trapping horizon, revealing
that their permissible vector fields would generally coincide with $\chi/L_\xi
R$. The current formulation in terms of $\chi$ is more manifestly invariant and
physically interpretable, with $\chi$ playing the role of a stationary Killing
vector.

The integrated flux due to gravitational radiation can similarly be written as
\begin{equation}
[E]_g\cong\int{*}\Theta(\chi,\tau)\wedge dx
\end{equation}
where the {\em effective gravitational-radiation energy tensor} $\Theta$ is
defined to have components which are determined as
\begin{eqnarray}
\Theta_{\pm\pm}&=&||\sigma_\pm||^2/32\pi\label{Theta0}\\
\Theta_{\pm\mp}&=&|\zeta_{(\pm)}|^2/8\pi e^f.\label{Theta1}
\end{eqnarray}
It has not been shown explicitly here that $\Theta$ is a tensor in the normal
space, since it is more easily seen in a spinorial formulation \cite{bhd4}.
Note that $\Theta$ is generally not symmetric; actually
$\Theta_{+-}\cong\Theta_{-+}$ with the admissible gauge choice $f\cong0$, but
the general expressions will be retained here, in view of future
generalizations. Since its components are non-negative, $\Theta$ satisfies the
dominant energy condition, implying that the gravitational radiation carries
positive energy. The four components of $\Theta$ are interpreted as the energy
densities of gravitational radiation: $\Theta_{++}$ for ingoing transverse
radiation, $\Theta_{-+}$ for ingoing longitudinal radiation, $\Theta_{+-}$ for
outgoing longitudinal radiation, and $\Theta_{--}$ for outgoing transverse
radiation, for $\theta_+\cong0$. Note also that $(\sigma_\pm,\zeta_{(\pm)})$
each have the correct number (two) of independent components for describing the
respective radiation. The identification and neat division of these four modes
is another success for the dual-null method.

The generator-volume form of the first law becomes
\begin{equation}
[E]\cong\int{*}(T(\chi,\tau)+\Theta(\chi,\tau))\wedge dx.\label{etg}
\end{equation}
In differential form,
\begin{equation}
L_\xi E\cong\oint{*}(T(\chi,\tau)+\Theta(\chi,\tau)).\label{etd}
\end{equation}
If the trapping horizon is spatial, one can use the unit normal vector
$\hat\tau=\tau/\sqrt{g_{xx}}$, recalling that
$g_{xx}=g(\xi,\xi)=-g(\tau,\tau)$, to give the proper-volume form
\begin{equation}
[E]\cong\int\tilde{*}(T(\chi,\hat\tau)+\Theta(\chi,\hat\tau))\label{etp}
\end{equation}
which is closest to that of Ashtekar \& Krishnan, having identified $\Theta$
and $\chi$. These three forms perhaps most clearly demonstrate the nature of
the first law as an energy-balance equation, expressing the increase in the
mass-energy $E$ of the black hole due to the energy densities of the infalling
matter and gravitational radiation.

\section{Surface gravity and a Gibbs-like equation}

In spherical symmetry, there is a natural definition of surface gravity
$\kappa$ for a trapping horizon, satisfying an equation with the same form as
the usual surface gravity for stationary black holes, but with the stationary
Killing vector replaced by the Kodama vector \cite{1st,in}. The energy-supply
term $\oint{*}\xi\cdot\psi$ in the first law can then be rewritten as $\kappa
L_\xi A/8\pi$, yielding an equation with the same form as the first law of
black-hole statics, which really is analogous to the Gibbs equation rather than
the first law of thermodynamics. In seeking a general definition of surface
gravity for trapping horizons, it is perhaps useful to note first that a formal
first law can be given for any energy density $w$, if the surface gravity is
related to it by
\begin{equation}
\oint{*}\kappa=4\pi\left(E-R\oint{*}w\right).\label{sg1}
\end{equation}
This yields the identity \cite{MH}
\begin{equation}
dE=\frac1{8\pi A}\oint{*}\kappa dA+\oint{*}wdR+Rd\left(\frac ER\right).
\end{equation}
Since $L_\xi(E/R)\cong0$ on a trapping horizon, projecting the above identity
along the trapping horizon yields
\begin{equation}
L_\xi E\cong\frac{1}{8\pi A}\oint{*}\kappa L_\xi A+\oint{*}wL_\xi R.
\label{gibbs1}
\end{equation}
This will be the desired {\em Gibbs-like equation for black holes},
generalizing that found in spherical symmetry \cite{1st}, with a similar form
to the first law of black-hole statics.

One wishes to define {\em surface gravity} $\kappa_{(\pm)}$ on a horizon with
$\theta_\pm\cong0$ and work density $w_{(\pm)}$ as above. This determines
$\kappa_{(\pm)}$ up to total divergences, and an argument in the next section
fixes it as
\begin{equation}
\kappa_{(\pm)}=-\frac R4e^f\left(2L_\mp\theta_\pm+\theta_+\theta_-\right).
\label{sg2}
\end{equation}
Here it is merely checked that $\kappa_{(\pm)}$ and $w_{(\pm)}$ are related as
above. The integrand in (\ref{sg1}) can be rewritten using the cross-focusing
equation (\ref{focus2}) as
\begin{equation}
e^f\left(2L_\mp\theta_\pm+\theta_+\theta_-\right)=e^f\left(16\pi
T_{+-}-\theta_+\theta_-\right)-\Re+2|\zeta_{(\pm)}|^2-2D\cdot\zeta_{(\pm)}^\sharp.
\end{equation}
The last term integrates to zero, while the second and third terms constitute
the integrand of the Hawking energy (\ref{energy}), integrating to $-16\pi
E/R$. Then
\begin{equation}
\oint{*}\kappa_{(\pm)} =4\pi E-4\pi
R\oint{*}\left(e^fT_{+-}+\frac{|\zeta_{(\pm)}|^2}{8\pi}\right)
=4\pi\left(E-R\oint{*}w_{(\pm)}\right)
\end{equation}
as claimed. A version of the Gibbs-like equation (\ref{gibbs1}) was given
previously in terms of the averaged surface gravity
$\oint{*}(\kappa_{(+)}+\kappa_{(-)})/2A$ and the averaged work density
$\oint{*}(w_{(+)}+w_{(-)})/2A$ \cite{MH}. The first law (\ref{first2}) resolves
such ambiguities by determining the appropriate $w$ and therefore $\kappa$. The
definition of outer trapping horizon \cite{bhd}, $L_\mp\theta_\pm<0$ (for
$\theta_\pm\cong0$), also indicates the suitability of $\kappa_{(\pm)}$ as
measures of surface gravity, since $\kappa_{(\pm)}>0$ for such horizons, with
$\kappa_{(\pm)}$ vanishing somewhere on degenerate horizons.

The standard definition of surface gravity for stationary black holes is via
the formula $\chi\cdot(\nabla\wedge\chi^\flat)\cong\kappa\chi^\flat$ on a
Killing horizon, where $\chi$ is the stationary Killing vector and a flat
($\flat$) denotes the covariant dual (index lowering). The same formula holds
for the preferred time vector $\chi$ on a trapping horizon, for an {\em average
surface gravity} $\bar\kappa$ defined below. One finds
$\chi^\flat=L_-Rdx^--L_+Rdx^+$ and
$2\nabla\wedge\chi^\flat=d\chi^\flat=2L_-L_+R\,dx^+\wedge dx^-$, noting that
$L_+L_-R=L_-L_+R$ since $R$ is an integral scalar of the transverse surfaces.
Then
\begin{equation}
\chi\cdot(\nabla\wedge\chi^\flat)=\bar\kappa dR
\end{equation}
where
\begin{equation}
\bar\kappa=-e^fL_-L_+R.
\end{equation}
In particular, on a trapping horizon with $\theta_\pm\cong0$,
$\chi^\flat\cong\pm dR$ and so
\begin{equation}
\chi\cdot(\nabla\wedge\chi^\flat)\cong\pm\bar\kappa\chi^\flat\label{asg}
\end{equation}
as desired. Thus $\bar\kappa$ is guaranteed to recover the standard expression
for stationary surface gravity if $\chi$ reduces to the stationary Killing
vector. If the expansions are constant on the transverse surfaces,
$D\theta_\pm=0$, then $\theta_\pm=2L_\pm R/R$, so that $\bar\kappa$ and
$\kappa_{(\pm)}$ all coincide. In particular, in spherical symmetry,
$\kappa=\bar\kappa$ recovers $\kappa\cong1/4m$ for Schwarzschild and
$\kappa\cong\sqrt{m^2-q^2}/(m+\sqrt{m^2-q^2})^2$ for Reissner-Nordstr\"om black
holes, where $m$ is the mass and $q$ the charge. The last result is
non-trivial, and has not been recovered by other definitions of surface gravity
for dynamical black holes.

For Kerr black holes, a dual-null foliation giving the correct $\chi$ and
$\kappa$ is not known, despite the recent construction of a pair of dual-null
foliations generating the horizons \cite{ker}. For the record, the results are
as follows, using tildes to denote quantities associated with those particular
dual-null foliations and reserving $\chi=\partial/\partial
t+\Omega\partial/\partial\phi$ and $\kappa=\sqrt{m^2-a^2}/2mr_+$ for the
correct values, where $m$ is the mass, $ma$ the angular momentum,
$(t,r,\theta,\phi)$ are Boyer-Lindquist coordinates, $r_+=m+\sqrt{m^2-a^2}$ is
the coordinate radius of the outer horizon and $\Omega=a/2mr_+$ is its angular
velocity. The results are $\tilde\chi=Q\chi$ and $\tilde\kappa=Q\kappa$ where
$Q=\tilde R^2(d\tilde R/dr)/\Sigma$, $\Sigma=r^2+a^2\cos^2\theta$ and $\tilde
R=(r^4+a^2r^2+2ma^2r)^{1/4}$. Thus, while $\tilde\chi$ and $\tilde\kappa$ are
out by a factor, it is the same factor. As a practical procedure, one could
rescale $\tilde\chi$ to $\chi$ to obtain the correct $\kappa$. It makes no
difference to use $\bar\kappa$, as $\tilde\theta_\pm=2\tilde L_\pm\tilde
R/\tilde R$ in this case. This is clearly less than satisfactory, though there
seems to be no other notion of surface gravity for dynamical black holes
without some ambiguity or apparently ad-hoc procedure to recover the Kerr
surface gravity. A guide here could be the issue of finding a dynamical zeroth
law, stating that the surface gravity becomes constant as a trapping horizon
becomes null. For $\bar\kappa$, one obtains $D\bar\kappa\cong0$ if $Df\cong0$,
which is a legitimate gauge choice on a trapping horizon of any signature. For
$\kappa_{(+)}$, the condition for $D\kappa_{(+)}\cong0$ is
$D(e^fL_-\theta_+)\cong0$, which constrains the null normal $l_-$ generating a
dual-null foliation away from a null trapping horizon. These issues are
currently unresolved.

\section{A Clausius-like equation, entropy flux and entropy conservation}

Given that stationary black holes theoretically possess an entropy $A/4$, the
parallel between the Gibbs-like equation (\ref{gibbs1}) and the thermodynamic
Gibbs equation (\ref{gibbs0}) suggest defining a {\em geometric entropy}
\begin{equation}
S\cong A/4
\end{equation}
for any trapping horizon \cite{HMA}. Units are such that the Planck and
Boltzmann constants are unity. The final issue to be addressed here is the
analogue of the second law of thermodynamics (\ref{second0}), which for a
non-isolated system states that the entropy of the system, minus the entropy
supplied to the system, is non-decreasing; that is, entropy is either produced
or conserved, but not destroyed. Then one needs a definition of geometric
entropy supply.

By comparing the first law (\ref{first2}) and the Gibbs-like equation
(\ref{gibbs1}), it was shown indirectly that
$\oint{*}\xi\cdot\psi\cong\oint{*}\kappa L_\xi A/8\pi A$. This is now shown
explicitly and the result localized. Combining
(\ref{flux2},\ref{focus1},\ref{horizon1},\ref{sg2}) successively yields
\begin{equation}
\xi\cdot\psi\cong\frac{\xi^\pm}{\xi^\mp}\frac{e^fL_\pm\theta_\pm}{8\pi}L_\xi
R\cong{}-\frac{e^fL_\mp\theta_\pm}{8\pi}L_\xi R\cong\frac{\kappa_{(\pm)}}{4\pi
R}L_\xi R
\end{equation}
and so
\begin{equation}
A\xi\cdot\psi\cong\frac{\kappa L_\xi A}{8\pi}.\label{flux3}
\end{equation}
Thus the energy flux through the horizon is unexpectedly proportional to the
surface gravity. Now recall as in \S II that heat flux $q$ is classically
proportional to temperature $\vartheta$, thereby defining entropy flux
$\varphi=q/\vartheta$, and that stationary black holes possess a Hawking
temperature $\vartheta=\kappa/2\pi$. Then the above result suggests defining a
{\em geometric entropy flux}
\begin{equation}
\varphi=\frac{2\pi\psi}\kappa.\label{entropy1}
\end{equation}
This argument is reminiscent of the original definition of entropy due to
Clausius, who argued that the heat supplied to a system, divided by
temperature, should be a total differential for closed cycles. Here $\varphi$
is not the classical thermodynamic entropy flux of the matter, but should be
regarded as a speculative definition of geometric entropy flux induced by
matter (or gravitational radiation) in a strong gravitational field. In other
words, if in some theory of quantum gravity one could indeed establish that
black holes have a geometric entropy $A/4$, perhaps one would also find that
infalling energy fluxes induce corresponding geometric entropy fluxes.

The corresponding {\em geometric entropy supply} $S_\circ$ (\ref{entropy0}) is
given by
\begin{equation}
L_\xi S_\circ=\oint{*}\xi\cdot\varphi\cong L_\xi A/4.
\end{equation}
The second law of thermodynamics (\ref{second0}) for trapping horizons would
state that $L_\xi S\ge L_\xi S_\circ$. However, we have $L_\xi S\cong L_\xi
A/4$ and so
\begin{equation}
L_\xi S\cong L_\xi S_\circ.\label{second1}
\end{equation}
Thus {\em geometric entropy is conserved}. This may sound radical, since what
is normally called the second law for black holes is an inequality, like the
second law of thermodynamics. A more faithful comparison with thermodynamics
has shown that, while the geometric entropy of a black hole generally
increases, it does so by the geometric entropy supplied to it by the infalling
matter and gravitational radiation, with no net entropy production. This might
perhaps be expected, since General Relativity is a classical theory which is
symmetric under time reversal. Entropy can be produced by quantum-mechanical or
statistical effects, in either the matter or quantum gravity, but appears to be
absent classically. Similarly, the generalized second law should be stated as
expressing combined (matter plus gravitational) entropy production, rather than
entropy increase.

\section{Conclusion}
The main results are summarized as follows. (i) An effective energy tensor
$\Theta$ (\ref{Theta0}--\ref{Theta1}) for gravitational radiation has been
identified for dynamical black holes. (ii) A Killing-like vector $\chi$
(\ref{chi}), providing a preferred flow of time outside a dynamical black hole,
has been identified and used to characterize the Ashtekar-Krishnan permissible
vector fields. (iii) The Ashtekar-Krishnan energy-balance equation for
dynamical black holes has been re-derived in a dual-null formalism, emphasizing
geometrical invariance and corroborating the physical interpretation as an
energy-balance law (\ref{etg}--\ref{etp}), with the black-hole mass-energy $E$
growing due to the energy densities of the infalling matter and gravitational
radiation, $T(\chi,\tau)$ and $\Theta(\chi,\tau)$. (iv) While the original
proper-volume form of the law applies only to spatial (dynamical) horizons,
here generator-volume and differential forms have been derived, which are both
regular in the physically important limit of null (isolated) horizons, where a
black hole is starved and ceases to grow. The new forms apply to any trapping
horizon, thereby describing inner black-hole horizons, white holes,
cosmological horizons, traversable wormhole mouths and evaporating black holes.
(v) The energy terms have been divided into those which vanish if and only if
the horizon is null, and those which generally do not (\ref{first2}), and
interpreted respectively as energy-supply and work terms, in analogy with the
first law of thermodynamics (\ref{first0}). (vi) A new definition of surface
gravity $\kappa$ (\ref{sg2}) has been given for dynamical black holes, such
that the energy supply can be written in terms of $\kappa$ and area $A$,
(\ref{flux3}), just as in the so-called first law for stationary black holes,
which is instead analogous to the Gibbs equation (\ref{gibbs0}) of
thermodynamics. An average surface gravity (\ref{asg}) has also been defined
with respect to $\chi$ by the same formula as stationary surface gravity. (vii)
Since the energy-flux covector $\psi$ is proportional to $\kappa$ on a trapping
horizon, the original Clausius concept of entropy suggests defining an entropy
flux $2\pi\psi/\kappa$ (\ref{entropy1}), and it follows that entropy is
conserved, (\ref{second1}), for dynamical black holes.

The last, perhaps surprising result does not contradict the fact that black
holes grow, either by Hawking's area theorem for event horizons
\cite{BCH,HE,MTW,Wal} or the area law for future outer trapping horizons
\cite{bhd}, which instead reflect the fact that, classically, a black hole is
the ultimate absorber. The black-hole area and presumed entropy $A/4$ increase,
but only by the entropy supplied to it by the infalling matter and
gravitational radiation. It should be acknowledged that this physical
interpretation is speculative, since it not known that dynamical black holes
have a truly thermodynamical entropy $A/4$ and a local temperature
$\kappa/2\pi$, in the same way as is known for stationary black holes. It
nevertheless illustrates that the classical first and second laws of black-hole
mechanics \cite{BCH,HE,MTW,Wal} are both misnomers. While versions of both laws
now exist for dynamical horizons, in addition, there are now more correct
analogues for black holes of the first and second laws of thermodynamics:
conservation laws for energy and entropy.

The most practical result here is probably (i), in the context of current
efforts to predict gravitational waveforms produced by dynamical black holes.
It has generally been believed that gravitational radiation is well defined
only in weak-field regimes, rather than the strong-field regime calculated by
numerical simulations, thereby posing the radiation-extraction problem: how to
extract the outgoing gravitational radiation and observable space-time strains
from the numerical simulations. Now, however, the energy tensor $\Theta$
defines gravitational radiation in the strong-field regime. Miraculously, a
dynamical black hole itself provides the required structure, the dual-null
foliation of ingoing and outgoing wavefronts generated from the trapping
horizon, as located by existing numerical methods. The gravitational radiation
may therefore be extracted from existing simulations by numerically
implementing the coordinate transformation to the preferred dual-null foliation
and calculating relevant quantities such as the conformal shear $\sigma_-/R$,
which yields the Bondi news \cite{mon,inf}, and the conformal energy flux
$R^2\psi_-$. The variables are simply related to the actual strain tensor
$\varepsilon/R$ to be measured by a gravitational-wave detector at large
distance $R$, via the conformal strain tensor
$\varepsilon=\int(\sigma_-/2R)dx^-$ \cite{gwbh}. It should be noted that
$\Theta$ is uniquely defined for a given trapping horizon. If one wishes to
have a definition which is unique for a given space-time, an appropriate
concept appears to be the trapping boundary \cite{bhd}, the boundary of an
inextendible region whose every point lies on some trapped surface. Under
smoothness assumptions, a trapping boundary is also a trapping horizon.
However, in practice, trapping horizons have exactly the right level of
uniqueness, since marginal surfaces are typically found numerically in any
reasonable slicing of a black-hole space-time. In an asymptotically flat
space-time, the waveforms must converge in advanced time and be equivalent for
different slicings, so the practical issue is whether the waveforms converge
sufficiently within the numerical domain; if so, it seems likely that they will
converge to equivalent waveforms for any reasonable slicing.

Strictly speaking, the results here have demonstrated the contribution of only
the $(\Theta_{\pm\pm},\Theta_{\pm\mp})$ components of the energy, with the
$(\Theta_{\mp\mp},\Theta_{\mp\pm})$ components determined by a symmetric
treatment of trapping horizons with $\theta_\pm\cong0$. However, the results
generalize to uniformly expanding flows of the Hawking energy away from a black
hole; the basic calculations may be found in \cite{mon}, with the formally
identical energy-tensor form to be reported subsequently
\cite{bhd4}\footnote{The author thanks Hugh Bray, Tom Ilmanen, Marc Mars,
Edward Malec and Walter Simon for discussions on these flows, which generalize
inverse mean-curvature flows. The latter, formulated as weak flows which can
jump over obstacles such as other black holes, were recently used to prove the
Riemannian Penrose inequality \cite{Bra,HI}, and the generalization provides a
possible route to the full Penrose inequality.}. An alternative (and probably
easier) numerical implementation would therefore be to locate the level
surfaces of these flows in the original spatial hypersurfaces, calculating
desired quantities separately for each hypersurface. The Hawking energy can
also be used to recover the Bondi energy loss at future null infinity
\cite{mon,inf}, which can also be written in a formally identical energy-tensor
form. Thus $\Theta$ apparently provides a good physical measure of the energy
densities of gravitational radiation all the way from a black-hole horizon out
to infinity. The radiation-extraction problem for dynamical black holes has
thereby been theoretically solved.

\bigskip\noindent
Acknowledgements. The work reported here was originally inspired by a
presentation of Abhay Ashtekar at the Erwin Schr\"odinger Institute in Vienna
during a workshop on the Penrose inequality. The author thanks the ESI for
local support and hospitality, and attendees for discussions. Partly supported
by the National Science Council of Taiwan under grant NSC 93-2811-M-008-043.

\end{document}